The proteomic to biology inference, a frequently overlooked concern in the interpretation of proteomic data: a plea for functional validation

Thierry Rabilloud 1, 2, 3*, and Pierre Lescuyer 4,5

1: CNRS UMR 5249, Laboratory of Chemistry and Biology of Metals, Grenoble, France

2: Univ. Grenoble Alpes, Laboratory of Chemistry and Biology of Metals, Grenoble, France

3: CEA Grenoble, iRTSV/CBM, Laboratory of Chemistry and Biology of Metals, Grenoble, France

4: Biomedical Proteomics Research Group, Department of Human Protein Sciences, Geneva University, Geneva CH-1211, Switzerland

5: Clinical Proteomics Laboratory, Department of Genetic and Laboratory Medicine, Geneva University Hospitals, Geneva CH-1211, Switzerland

*: to whom correspondence should be addressed:

Laboratoire de Chimie et Biologie des Métaux, UMR CNRS-CEA-UJF 5249, iRTSV/LCBM, CEA Grenoble, 17 rue des martyrs, F-38054 Grenoble Cedex 9, France
thierry.rabilloud@cea.fr

Abstract

Proteomics will celebrate its 20th year in 2014. In this relatively short period of time, its has invaded most areas of biology and its use will probably continue to spread in the future. These two decades have seen a considerable increase in the speed and sensitivity of protein identification and characterization, even from complex samples. Indeed, what was a challenge twenty years ago is now little more than a daily routine. Although not completely over, the technological challenge now makes room to another challenge, which is the best possible appraisal and exploitation of proteomic data to draw the best possible conclusions on a biological point of view. The point developed in this paper is that proteomic data are almost always fragmentary. This means in turn that although better than a mRNA level, a protein level is often insufficient to draw a valid conclusion on a biological point of view, especially in a world where post-translational modifications play such an important role. This means in turn that transformation of proteomic data into biological data requires an important intermediate layer of functional validation, i.e. not merely the confirmation of protein abundance changes by other methods, but a functional appraisal of the biological consequences of the protein level changes highlighted by the proteomic screens.

Many if not most proteomic studies aim at comparing at least two different biological states to find differences in protein expression, most often quantitative differences. The conceptual rationale of these studies is based on homeostasis at the cell and organism levels, so that the proteins that change are perceived as important and linked to the differences in phenotypes between the various states compared. This concept is used by all functional -omics and is true for physiology or fundamental cell biology. In a nutshell, the concept is that changes observed with -omics translate into biologically relevant events or mechanisms.

When proteomics is applied as the discovery tool for such comparative purposes, two major hurdles must be overcome.

The first hurdle is the one of protein inference, i.e. how to convert mass spectrometry (MS) signals acquired on digestion peptides, or, in the case of gel-based proteomics, a combination of MS signals and image intensities, into a list of modulated proteins. MS data processing to obtain trustful lists of identified and relatively quantified proteins has been the subject of intense debate and investigation and was at the origin of proteomics data publication guidelines already almost 10 years ago [1-3]. As a positive consequence of those guidelines, no proteomics dataset can today be published without false discovery rate estimation at the peptide level, although the peptide to protein inference problem can still not be considered as a straightforward issue [4]. Concerning quantitative results extraction from MS data, the guidelines are still evolving, obviously as the field is not as mature as MS/MS identification interpretation. Overall, those guidelines have participated in significantly increasing the quality of published proteomics datasets, especially in the handful of specialized proteomics journals. Although not fully settled, this first hurdle will not be further commented in this paper.

The second hurdle is the protein to biology inference and it is felt that the problems linked to this step are often overlooked, leading possibly to data over- or misinterpretation and thus to misleading conclusions. We would therefore like to comment on this biology inference problem and suggest some guidance, based on published examples, to circumvent possible interpretation issues.

Although shotgun proteomics and gel-based proteomics behave differently in the protein to biology inference problem, the common root shared by all proteomic setups is that the characterization of the protein changes is most often not sufficient to draw solid biological conclusions directly. First, protein identification is based on subsets of peptides and information for correct characterization of protein family members or protein isoforms (PTM, cleavage) may be missing. Second, proteomic analysis rarely identifies all protagonists from a given metabolic pathway or functional network, and information on modulating or limiting factors is often lacking. Third, even assuming reliable identification and quantification, measuring a change in protein level does not always tell much about the biological effect, particularly when considering that many proteins have multiple functions or exist as active and inactive isoforms. In all these situations, the problem is the same: how to draw reliable conclusions on biological processes from sketchy and incomplete data? This should obviously not been considered as an irreducible flaw, and proteomics offers great chances for solving biological and biochemical questions. It rather means that the strengths and limitations of the toolbox must be clearly understood to obtain the greatest solid benefit.

In the case of shotgun proteomics, the most frequent situation is that proteins are identified by only a few peptides, covering a very low and non-uniform portion of the protein sequence. From these few peptides, a change in the biological activity represented by the protein is generally inferred, but this can be a strong overinterpretation.
One reason is that the overall length of the protein is not documented. It is very rare that the identified peptides are positioned close to the termini of the proteins, so that it is not possible to know whether the complete protein or only a fragment is detected. This is trivial in gel-free proteomics, where only peptides are analyzed without any prior separation at the protein level. This is however also true in the GeLC setup, where SDS-PAGE of proteins is used as the first separation stage, followed by in-gel digestion and LC/MS/MS of the peptides [5]. This is due to the fact that the resolving power of SDS PAGE is not fully used, as a limited number of bands are usually cut in a shortly migrated gel lane. Just to take an example, if a gel able to separate proteins between 20 and 200 kDa is cut in ten fractions, the ten fractions will cover the following Mw ranges from 20-25 kDa to 159-200 kDa, through windows such as 40-50kDa or 80-100kDa. This is due to the logarithmic law governing the relationship between migration distance and molecular mass [6]. The consequence is a 20% inaccuracy in the protein mass determination for each fraction, i.e. an important room for error. It has been shown that forms differing even moderately in molecular mass can have quite different properties, e.g. on an immunological or clinical point of view [7, 8]. Protein cleavage represents indeed a very complex way of regulation of protein activity, as exemplified by trypsin itself. After zymogen activation, which releases the active protease, trypsin autolyses progressively, producing various cleavage products with trypsin activity [9, 10], no proteolytic activity at all [11], or with no trypsin activity but a low chymotrypsin activity [12]. This example underlines the difficulty to infer the actual function of a protein from only a small subset of peptides, and the tendency to neglect the molecular mass information [13] is likely to introduce further difficulties in the functional interpretation of the proteomic data.

More generally, it is now well-recognized but still underestimated that PTM represent a very important way of modulating protein activity, and this statement goes much beyond the classical example of phosphorylation. Acetylation [14, 15], methylation [16], glycosylation [17], or prenylation [18] have been documented to modulate strongly protein localization and/or activity. Even minimal chemical modifications, such as oxidation, can result in protein inactivation [19-21]. Thus, the absence of knowledge of the precise modification profile of a protein can result in a misleading interpretation of the observed quantitative changes. Indeed, an increase in a protein amount can reflect not a boost in the protein activity, but just an attempt of the cells to maintain a biological function that has been altered, e.g. by an oxidation (e.g. in [22]). Conversely, an activating PTM occurring on a protein (e.g. by phosphorylation or acetylation) can boost a protein function at constant level [14, 15] and may even compensate a decrease in protein level.

It can be argued that the documentation provided in shotgun proteomics is comparable to the indirect one given by transcriptomics, or to the one obtained through the epitopes recognized in a sandwich ELISA. This is certainly true, but it should be kept in mind that ELISA are rarely used as discovery tools but rather as scoring tools, and that transcriptomic data are now not considered in most journals as a sufficient evidence by themselves. Put bluntly, if not provocatively, the limited characterization afforded by shotgun proteomics, when used in a generic discovery mode, does not usually offer by itself sufficient information to infer a change in the protein activity, and thus to correlate reliably the proteomic data to biology.

This caveat does not apply only to proteomics but to most omics, and just reflects an often overlooked lesson of genomics, i.e. that complex genomes are not that complex in terms of range of gene products expressed. For example, the human genome contains ca. 21,000 protein-coding genes [23], to produce an organism with ca. ten thousand billion cells [24]. This is only 10% more protein-coding genes than the worm genome, producing an organism with 1000 cells [25], and less than four times the number of protein-coding genes of the unicellular yeast genome [26]. This means in turn that the increase in regulations complexity is likely to hide not in the number of gene products, but in combinatorial regulations downstream, i.e. at the protein interaction level and at the PTM combinatorial level, both being of course intertwined. These downstream layers of regulations can produce paradoxical results when compared to mRNA or protein levels alone. This can be due to various modifications, including protein cleavages as mentioned above, but also to stoichiometry issues in protein complexes. As an example, when present in excess, a scaffolding subunit in a multiprotein complex can lead to a decreased activity by decreasing the proportion of fully assembled and fully active complexes. Thus, it should be kept in mind that a direct inference from a protein level to a biological effect can be an overspeed conclusion.

Although different, the situation in 2D gel-based proteomics is problematic as well. 2D gels provide a rather accurate characterization of the molecular mass and pI of the protein of interest, and will allow determining whether it is the full-length protein or a fragment that is investigated. Moreover, multiple PTM often result into multiple spots [27], so that it is often possible to know which form of the protein is changing. Knowing the exact nature and position of the PTM can however represent a difficult task [21, 28], at least in bottom-up proteomics. Last but not least, 2D gel-based proteomics also shows the unique ability to compare protein abundances within one sample due to the uniformity of protein staining [29]. This said, the main danger of overinterpretation in 2D gel-based proteomics is embedded in the way 2D gels are used to select quantitative changes. Most often, changes are detected on a spot by spot basis, so that proteins appearing as multiple spots pose the problem of taking a part for the whole. A typical example is malate dehydrogenase, which appears as three spots (Figure 1). In some circumstances, only the most acidic (i.e. most modified form) is induced, the other major spots being constant. In the classical 2D gel-based proteomic setup, the increased spot is excised, its nature determined by mass spectrometry, and an increase in malate dehydrogenase is deduced. On a purely quantitative point of view, this deduction is a nonsense, as the minor spot represents only a few percent of the total malate dehydrogenase. It happens however that malate dehydrogenase activity is strongly modulated by PTM such as acetylation [14], so that this minor spot may bear an important part of the malate dehydrogenase activity and its change may be relevant, functionally speaking. In this context, it is impossible to know which is the real functional inference to be made on the sole basis of the proteomic data.

In conclusion, no discovery proteomic setup is devoid of interpretation biases at the stage of drawing biological inference from proteomic data. Various parameters, such as the addressed biological question, study design, sample type and processing, will have an impact on data interpretation and must be carefully defined. Nevertheless, and independently of these aspects, validation by independent and preferably orthogonal techniques appears as an unavoidable process to consolidate the proteomic data.

This validation can be divided in two distinct parts. The first stage of validation deals with the problem of protein inference and fragmentary data and aims at confirming the identity of the protein species/isoform/PTM identified from shotgun experiments or 2D gel-based proteomics. It will also confirm quantitative data from proteomic experiments. This first stage of verification is usually performed with independent methods like immunoblotting (e.g. in [30] ) , qPCR, or SRM (e.g. in  [31] ).

Although important for securing the protein identifications and the quantitative changes observed, this validation at the sole protein level is often not sufficient to infer biological changes. A second stage of validation is requested to address the question of functional inference. In most cases, what is inferred from the protein quantitative differences is a change in the activity of the protein.  Due to the numerous caveats discussed above, this has to be experimentally verified, and the functional validation of proteomic data can take various forms. The objective of this article is clearly not to review all of them, but to provide some examples, starting with siRNA, which can be extremely a valuable tool for investigating cell biology processes [32]. In many cases, however, a change in activity can be directly or indirectly documented by a biochemical validation. This can be an enzyme assay, a metabolite assay [30], or an indirect assay based on pharmacological inhibitors [33].

The example of the pyruvate assay used in the frame of schizophrenia by Martins-de-Souza et al. also illustrates an important point, which is that specificity should not be mistaken for relevance. Since the description of the "déjà vu in proteomics" [34, 35], there is a tendency to consider that changes in metabolic enzymes represent a general adaptation of cells that has little to do with relevant cellular responses. This lecture grid is strongly influenced by the biomarker perspective, where specificity and selectivity are essential. Although not specific, changes in metabolic enzymes can be critical on a mechanistic point of view. Impaired carbohydrate metabolism in schizophrenia is a good example [36, 37], as well as sterol metabolism for immune control and inflammatory diseases [38, 39]. In this frame, the added value of proteomics is both to point out the pathway of interest (and thus pharmacological entry points to modulate it) and the proteins modulated within the pathway (e. g. in [40, 41] for statins).

The latter example points out to another trend in proteomics that has developed due to the specific features of shotgun proteomics. In general, shotgun proteomics (or transcriptomics) delivers a huge list of modulated proteins, generated by automated *in silico* database search, so that individual validation becomes impossible. Researchers usually resort to pathway analysis to make the most of such massive data. These tools are helpful since they can provide some hints for further investigation of some of the identified proteins. However, the too frequent weak point is that the outcome of pathway analyses is not validated by additional experiments, so that the reader is left short of solid data. There are happy exceptions however, as shown by an example on the proteomic analysis of Cornelia de Lange syndrome [42]. This study highlighted a variety of proteins with no obvious link between themselves, but on the one hand oxidative stress was suspected and on the other hand the pathway analysis converged to Rad21 and c-myc. The authors validated both proteomic-driven hypotheses by a combination of western blotting, chromatin immunoprecipitation and qPCR. By doing so they closed the circle on both hypotheses and ended up with solid data that can be convincing much beyond the proteomic community.

In conclusion, proteomic data are often not precise enough to make valid biologically inferences directly. Thus, targeted validation by non proteomic methods seems indispensable to warrant the validity and functional importance of the findings made by proteomics. It seems to us that ten years after securing the quality of the MS data, tending to this goal of securing the biological relevance should be the next objective for the proteomic literature.

The authors have declared no conflict of interest

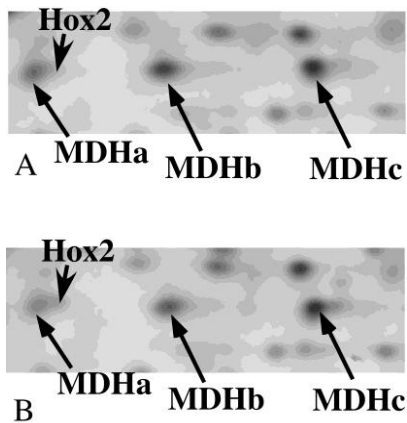

Figure 1: RAW 264 macrophage cells were cultured either under control conditions or treated for 24 hours with 0.11mM zinc sulfate. The cellular proteins were extracted in a urea-thiourea buffer, and separated by 2D electrophoresis (1st dimension linear pH gradient from 4 to 8, 2nd dimension 10% acrylamide gel). Only the zones containing the malate dehydrogenase (Swissprot P14152) spots are shown.
When the control (Panel A) and zinc-treated (panel B) conditions are compared by computerized image analysis (Delta 2D software) a significant change (reduction by zinc treatment) is found in spot a, but not in spots b and c, nor in the Hox2 spot.